\documentclass[letters, fleqn, usenatbib]{mnras}
\usepackage{amssymb,amsmath}
\usepackage{graphicx}
\usepackage{xcolor}
\usepackage{multirow}
\usepackage[T1]{fontenc}
\usepackage{ae,aecompl}
\usepackage{newtxtext,newtxmath}

\def\msig{$M_{\rm BH}- \sigma$\ }

\title[Outlier in \msig space]
{SDSS J0159 as an outlier in the \msig space: further clues to 
support a central tidal disruption event?}
\author[Zhang, Bao \& Yuan]
{Xue-Guang Zhang\thanks{Corresponding author
    Email: \href{mailto:xgzhang@njnu.edu.cn}{xgzhang@njnu.edu.cn}}, Min Bao, QiRong Yuan\\
    School of physics and technology, Nanjing Normal University,
                No. 1, Wenyuan Road, 210046, P. R. China}
\date{}

\begin{document}
\pagerange{\pageref{firstpage}--\pageref{lastpage}} \pubyear{2019}
\maketitle
\label{firstpage}

\begin{abstract}
In this Letter, properties of black hole (BH) mass are well checked 
for the interesting object SDSS J0159, a changing-look AGN 
and also a host galaxy of a tidal disruption event (TDE). Through 
spectral absorption features, the stellar velocity dispersion 
of SDSS J0159 can be well measured as $\sigma\sim81~{\rm km/s}$, 
leading to SDSS J0159 being an apparent outlier in the \msig space,  
because of the BH mass estimated through the \msig relation about two 
magnitudes lower than the reported virial BH mass of about 
$10^8~{\rm M_\odot}$. After considerations of contributions of stellar 
debris from the central TDE to broad line emission clouds, the 
over-estimated virial BH mass could be well explained in SDSS J0159. 
Therefore, over-estimated virial BH masses through broad line 
properties in the \msig space could be treated as interesting 
clues to support central TDEs.
\end{abstract}

\begin{keywords}
galaxies:active - galaxies:nuclei - quasars:emission lines - galaxies:Seyfert
\end{keywords}

\section{Introduction}
 
      SDSS J0159 (=SDSS J015957.64+003310.5) is an interesting 
object, not only due to its being firstly classified as a 
changing-look QSO in \citet{ld15} with its apparent type transition 
from a type 1 quasar in 2000 to a type 1.9 Active Galactic Nuclei 
(AGN) in 2010, but also due to its long-term variabilities over 
ten years well explained by a tidal disruption event (TDE) 
in \citet{md15}.

     The nature of changing-look AGNs (CLAGNs) is not totally 
clear so far. More and more evidence strongly indicates that 
dust obscuration should be not the main reasons, but variations 
of accretion flows should play the key roles as well discussed 
in \citet{ld15} and more recently discussed in \citet{yw18}. Once 
variations of accretion flows were accepted to CLAGNs, there 
should be two scenarios. 

	On the one hand, variations of accretion flows have few 
effects on dynamical structures of broad emission line regions 
(BLRs). For this kind of CLAGNs, it should be well confirmed that 
dynamical properties of broad emission line clouds moving in 
Keplerian orbits lead to reliable virial BH masses estimated 
by properties of observed broad emission lines under the 
virialization assumption \citep{pf04, rh11} combining with the 
efficient empirical R-L relation applied to estimate BLRs sizes 
through continuum/broad line luminosity \citep{kas05, bd13}. 

	On the other hand, variations of accretion flows 
have apparent effects on dynamical structures of BLRs, 
such as those cases related to well-known TDEs \citep{re88, 
ve11, ce12, gs12, gr13, gm14, ko15, ht16, wy18}. 
Based on the more recent simulated results on TDEs in 
\citet{gr13, gm14}, accreting fallback debris could provide 
efficient materials leading to observed broad emission lines, 
however, there are apparent time-dependent structure evolutions 
of BLRs from TDE debris. In other words, the empirical R-L 
relation is not efficient to estimate sizes of the BLRs built 
by debris from TDEs to some extent, such as the reported abnormal 
properties of time-dependent variabilities of broad emission 
lines in the TDE discussed in \citet{ht16}. Therefore, due 
to contributions of TDE debris, more effects should be 
considered to determine virial BH mass by properties of 
broad line width and line (or continuum) luminosity. 

    As a CLAGN and also a TDE candidate for SDSS J0159, it is 
interesting to check whether are there effects of the probable 
central TDE on properties of the virial BH mass through broad 
lines. The main objective of the Letter is trying to find 
probable clues to support or go against the central TDE 
in SDSS J0159, after considerations of probable contributions 
of TDEs to broad emission lines. Section 2 presents our main 
results on BH mass properties of SDSS J0159, and necessary 
discussions. Section 3 gives our final conclusions.

\section{BH mass of SDSS J0159}
\subsection{Virial BH mass} 
  
   Under the widely accepted virialization assumption 
for broad emission lines in SDSS J0159, \citet{ld15} have 
estimated virial BH mass of $(1.7\pm0.1)\times10^8~{\rm M_\odot}$ 
and $(1.6\pm0.4)\times10^8~{\rm M_\odot}$ by properties of the 
single-epoch spectrum in 2000 and 2010, respectively. Meanwhile, 
\citet{md15} have reported virial BH mass of 
$1.3~\sim~1.6\times10^8~{\rm M_\odot}$, similar as the 
values in \citet{ld15}.

   Before proceeding further, there is one point we should note. 
As the results in \citet{ld15}, it is clear that there are strong 
effects of obscuration on broad Balmer emission lines in 2010, 
due to the non-detected broad H$\beta$ but apparent broad H$\alpha$ 
in the spectrum. As a consequent result, the applied luminosity 
of broad H$\alpha$ was underestimated in 2010, leading to lower 
virial BH masses of SDSS J0159 estimated through spectrum 
observed in 2010. Therefore, values from spectrum in 2000 are 
mainly considered.

     Moreover, besides the formulas discussed in \citet{vp06, gh07, 
gr10} accepted by \citet{ld15} and \citet{md15}, based on the 
spectroscopic features in 2000, the second moment of broad H$\beta$ 
(about 2098{\rm km/s}) combining with the BLRs size of about 
30${\rm light-days}$ determined by the empirical relation in \citet{bd13} 
has been applied to estimate the virial BH mass by the formula 
well discussed in \citet{pf04} 
$M_{\rm BH} = 5.5 \times\frac{\sigma(H\beta)^2\times R_{BLRs}}{G} 
=1.6\times10^8{\rm M_\odot}$, well coincident with the ones in 
\citet{md15} and in \citet{ld15}. Therefore,
$M_{\rm BH}~=~(1.7\pm0.1)\times10^8~{\rm M_\odot}$ was safely
accepted as the virial BH mass of SDSS J0159 in the letter.

\begin{figure*}
\centering\includegraphics[width = 18cm,height=5cm]{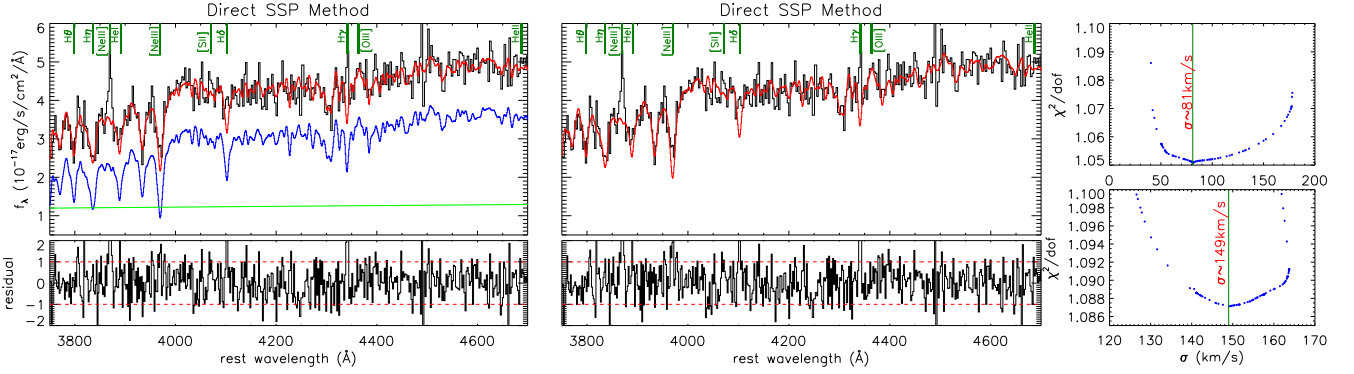}
\caption{The best fitted results and the corresponding residuals 
to the spectral absorption features by the direct SSP method with 
(left panels) and without (middle panels) considerations of a power law 
component. In top left and middle panels, solid lines in black and in 
red represent the observed spectrum and the best fitted results, 
respectively. Thick solid lines in dark green mark 
the emission lines being masked out when SSP method applied.  
In top-left panel, solid lines in blue and in 
green show the determined stellar component and the determined power 
law component, respectively. In bottom left and middle panels, solid 
lines in black show the corresponding residuals, and dashed lines 
in red show $residual=\pm1$. Right panels show 
the dependence of $\chi^2/dof$ on stellar velocity dispersions for model 
functions with (top right panel) and without (bottom right panel) a 
power law component considered.
}
\label{ssp}
\end{figure*}

\subsection{Stellar velocity dispersion of SDSS J0159}

\begin{figure}
\centering\includegraphics[width = 8cm,height=5cm]{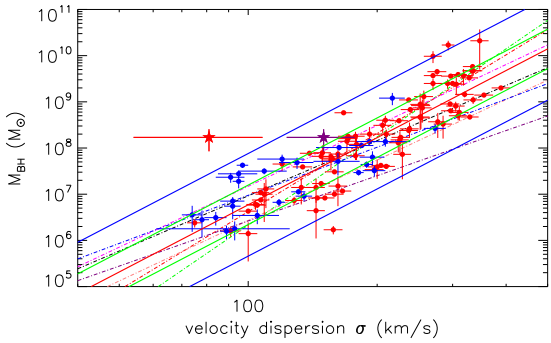}
\caption{On the correlation between stellar velocity dispersions and 
BH masses. Solid five-point-star in red shows the 
values of SDSS J0159 with the best estimate of $\sigma\sim81{\rm km/s}$, 
and solid five-point-star in purple shows the values of SDSS J0159 
with $\sigma\sim150{\rm km/s}$. Solid circles in red and in blue 
show the values for the 89 quiescent galaxies and the 29 RM AGNs, 
respectively. Solid lines in red, in green and in blue show the 
best fitted results and the corresponding 68\% and 99\% confidence 
bands, respectively. Dot-dashed lines in green, in red, in magenta, 
in black, in pink, in purple and in blue represent the \msig relations 
listed from top to bottom in Table~1.}
\label{msig}
\end{figure}

   Besides the virialization method, the well-known \msig relation 
can also be well applied to estimate central BH masses of both 
quiescent galaxies and AGNs. The \msig relation was firstly reported 
by \citet*{fm00, ge00}, based on dynamic measured BH masses 
and measured stellar velocity dispersions of a small sample 
of nearby quiescent galaxies. More recent review of the \msig 
relation for galaxies can be found in \citet{kh13}. And 
\citet{mm13} and \citet{sg15} have shown more recent and detailed 
results on the \msig relations for large samples of quiescent 
galaxies with dynamic measured BH masses. Meanwhile, many studies 
have reported applications of the \msig relation from quiescent 
galaxies to AGNs \citep*{fg01, wt06, gu09, hk14, wy15}. 
More recent and detailed discussions on the \msig relations 
to both AGNs and quiescent galaxies can be found in 
\citet{bt15}: the \msig relation for broad line AGNs has the same 
intercept and scatter as that of reverberation mapped AGNs as well as 
that of quiescent galaxies. Here, in order to estimate BH mass 
of SDSS J0159 through the well-improved \msig relation, the stellar 
velocity dispersion should be firstly measured.

   There are two methods applied to measure the stellar velocity 
dispersion of SDSS J0159, based on Simple Stellar Population 
method (SSP method) applied with a bit different techniques in 
details: the direct SSP method and the STARLIGHT code. Here, due 
to higher quality of spectral absorption features, the spectrum 
within rest wavelength from 3750\AA~ to 4700\AA~ observed in 2010 
(PLATE-MJD-FIBERID~=~3609-55201-0524) rather than in 2010 is 
mainly considered, because of none-detected absorption features 
around Mg~{\sc i}$\lambda5175$\AA~ and Ca~{\sc ii}$\lambda8498, 8542, 8662$\AA\ 
triplet. Moreover, due to similar measured stellar velocity 
dispersions through absorption features in different wavelength regions 
\citep*{ba02, gh06, xb11, wy15}, there should be no discussions on 
effects of different wavelength regions on measured stellar velocity 
dispersions.

  Detailed descriptions on the direct SSP method can be found 
in \citet{bc03, ka03}. Based on the 39 broadened simple stellar 
population template spectra in \citet{bc03} plus one power law 
function applied to describe AGN continuum emissions, the observed 
spectrum of SDSS J0159 can be well described through the 
Levenberg-Marquardt least-squares minimization technique 
(the $MPFIT$ package). Certainly, when SSP method applied, 
emission lines have been masked out with FWZI (Full Width at Zero 
Intensity) about 400${\rm km/s}$, which are marked by dark green 
lines in Fig.~\ref{ssp}. Left panels of Fig.~\ref{ssp} show the 
best fitted results with $\chi^2/dof~=951.3/892~=1.07$ (total 
squared residuals divided by degree of freedom) and the 
corresponding residuals $\frac{y-y_{m}}{y_{e}}$ with $y$, 
$y_{e}$ and $y_{m}$ represent the observed spectrum, the 
corresponding errors and the best fitted results, respectively. 
Top-right panel shows the sensitive dependence of $\chi^2/dof$ 
on stellar velocity dispersions, as strong evidence to support 
reliable measured stellar velocity dispersion. And the determined 
stellar velocity dispersion is about $\sigma = (81.4\pm26.7)~{\rm km/s}$. 

   Before proceeding further, there is one point we should note. 
Through the spectral features in 2010, SDSS J0159 has been 
classified as a main galaxy in SDSS with a pipeline measured 
stellar velocity dispersion of $\sigma_{\rm SDSS}\sim169\pm24~{\rm km/s}$, 
much larger than our measured value. The main reason leading 
to the different stellar velocity dispersions is as follows. 
In our code of the direct SSP method, there is one additional 
power law component, besides the stellar template spectra. Once 
the additional power law component was removed from the templates, 
only the stellar template spectra were applied with the direct 
SSP method, leading to $\chi_2^2/dof_2~=971.9/894~=1.09$ and 
$\sigma=149.9\pm22.8~{\rm km/s}$ which is similar as the SDSS 
reported value. The best fitted results and the corresponding 
residuals are shown in middle panels of Fig.~\ref{ssp}, 
and the dependence of $\chi_2^2/dof_2$ on $\sigma$ is shown in 
the bottom-right panel. In other words, there are two models 
with different applied model functions leading to the different 
determined stellar velocity dispersions. Therefore, it is 
necessary to check which model is more appropriate, through 
the F-test technique applied as follows.

   Based on the best fitted results by the two models, the 
F-value is firstly calculated as 
\begin{equation}
	F = \frac{(\chi_2^2 - \chi^2)/(dof_2-dof)}{\chi^2/dof} = 9.66
\end{equation},
where $\chi_2^2=971.9$, $dof_2=894$ and $\chi^2=951.3$, $dof=892$ 
represent the sum of squared residuals and the corresponding 
degrees of freedom for the best fitted results by the model 
functions without a power law component considered and with 
the power law component considered, respectively. Through 
the comparison between the calculated F-value $F=9.66$ and 
the F-value estimated by the F-distribution with the numerator 
degrees of freedom of $dof_2-dof$ and the denominator degrees 
of freedom of $dof$, we can conclude which model is preferred. 
The F-value by the F-distribution with p=0.0001 is 9.31 (the 
IDL function f\_cvf(0.0001, 2, 892)) based on the numerator 
and denominator degrees of freedom, which is smaller than 
$F=9.66$. The result strongly indicates that it is more 
preferred to describe the spectral absorption features by 
the stellar template spectra plus the power law component, 
with confidence level higher than 1-p=99.99\%. Meanwhile, the 
observed broad H$\alpha$ also indicates continuum emissions 
included in the observed spectrum in 2010, such as the 
determined power law continuum emissions in \citet{ld15}. 
Therefore, we finally accepted 
$\sigma=(81.4\pm26.7)~{\rm km/s}$ as the best estimate to the 
stellar velocity dispersion of SDSS J0159 through the direct SSP 
method, even though the spectrum has lower spectral quality.

   Besides the direct SSP method simply applied above, 
the widely accepted STARLIGHT code has been applied to 
measure stellar velocity dispersion of SDSS J0159 with more 
stellar template spectra, in order to ensure our measured 
stellar velocity dispersion is reliable. The more detailed 
descriptions on STARLIGHT code can be found in \citet{cf05}. 
Based on the 159 stellar template spectra created through 
\citet{bc03}, the estimated stellar velocity dispersions are 
about $\sigma\sim79~{\rm km/s}$ and $\sigma\sim151~{\rm km/s}$ 
by the STARLIGHT code applied with and without a power law 
continuum component considered, respectively. The results are 
much similar as the ones through the direct SSP method above. 
And we do not show further results on the best fitted 
results to the absorption features in SDSS J0159 by the 
STARLIGHT code any more, which are totally the same as the 
ones in Fig.~\ref{ssp}. Moreover, after considerations 
of the power law component, based on the 
STARLIGHT code, the calculated total stellar mass is about 
$7.1\times10^{10}{\rm M_\odot}$ very similar as the one
reported in \citet{md15}. Therefore, with full 
considerations of existence of the power law continuum 
emissions, STARLIGHT code also leads to $\sigma\sim80~{\rm km/s}$ 
of SDSS J0159. 

    The two different methods lead to the similar stellar velocity 
dispersion after considerations of the truely existed power law continuum 
emissions in the spectrum, indicating the measured value should be 
reliable to some extent. Certainly, due to lower 
spectral quality,the velocity dispersion is more likely to be around 
81{\rm km/s} as the best estimate, but the larger dispersion may not 
be absolutely ruled out.

\subsection{Properties of SDSS J0159 in the \msig space}

   Based on the measured stellar velocity dispersion and the 
reported virial BH mass of SDSS J0159, it is interesting to 
check properties of SDSS J0159 in the plane of stellar velocity 
dispersions and BH masses (the \msig space). 

   Before proceeding further, there are two points we should note. 
On the one hand, after considerations of reliability of BH masses, 
we only consider the quiescent galaxies with reported BH masses 
estimated/measured by stellar/gas dynamical techniques and the 
reverberation mapped AGNs (RM AGNs) with reported BH masses by 
emission line variability properties from multi-epoch spectra. The 
considered quiescent galaxies and the RM AGNs are listed in Table 1. 
There are finally 89 quiescent galaxies (including the three 
galaxies with BH masses larger than ${\rm\sim10^{10}~M_\odot}$ 
in \citet{mm11} and in \citet{tm16}) and 29 RM AGNs collected 
from the literature with more reliable BH masses.

   On the other hand, through different samples of active and/or 
quiescent galaxies, there are much different \msig relations, 
\begin{equation}
	\log(\frac{M_{\rm BH}}{\rm M_\odot})~=~\alpha~+~\beta~\times~
	\log(\frac{\sigma}{\rm 200~km/s})
\end{equation},
reported in the literature, such as the results listed in Table 1. 
Here, after considerations of uncertainties of both BH masses and 
stellar velocity dispersions for the collected both quiescent 
galaxies and RM AGNs with more reliable BH masses, the \msig relation 
is re-calculated as, 
\begin{equation}
	\log(\frac{M_{\rm BH}}{\rm M_\odot}) = (8.22\pm0.04) + 
	(4.81\pm0.21)\times\log(\frac{\sigma}{200~{\rm km/s}})
\end{equation}.
Here, the Least Trimmed Squares (LTS) regression technique discussed 
in \citet{ca13} is applied to find the best fitted results and  
the corresponding 68\% and 99\% confidence bands. The results are 
shown in Fig.~\ref{msig}

   Now, it is interesting to check properties of SDSS J0159 in the 
\msig space, shown in Fig.~\ref{msig}. It is clear 
that SDSS J0159 is an apparent outlier in the space if the best 
estimate of $\sigma\sim81{\rm km/s}$ accepted, and the virial 
BH mass is about two magnitudes higher than the value estimated 
through the \msig relation. Unfortunately, once $\sigma\sim150{\rm km/s}$ 
accepted, SDSS J0159 in the space also deviates from the best 
description on the \msig relation, but is not an apparent outlier 
in the space. Here, in the letter, as well discussed results above 
on the measured stellar velocity dispersions, the following discussions 
on SDSS J1059 are based on the point that the best estimate of 
stellar velocity dispersion is about $81{\rm km/s}$. And moreover, 
even accepted the discussed results in \citet{hk14}: BH masses 
of objects with pseudo-bulges could be statistically lower 
than the values estimated by the \msig relation, it is hard 
to explain why SDSS J0159 is an outlier in the \msig space, 
because SDSS J0159 has its virial BH mass higher not lower 
than the value estimated through the \msig relation. 
Therefore, properties of pseudo-bulge can not be applied 
to explain why is SDSS J0159 an outlier in the \msig space, 
and no further discussions on properties of bulge of SDSS J0159.

\begin{table}
	\caption{The accepted \msig relations}
\begin{tabular}{lcccc}
\hline\hline
obj    &   N  &   $\alpha$  & $\beta$  & Ref \\
\hline
QG  &  89  & 8.24$\pm$0.10 & 6.34$\pm$0.80 &
       \citet{sg15} \\
QG  &  72  & 8.32$\pm$0.05  & 5.64$\pm$0.32 &
       \citet{mm13} \\
QG  & 51  &  8.49$\pm$0.05  & 4.38$\pm$0.29 &
        \citet{kh13} \\
RA  & 29 & 8.16$\pm$0.18 & 3.97$\pm$0.56 &
        \citet{wy15} \\
RAC & 16 & 7.74$\pm$0.13 & 4.35$\pm$0.58 &
       \citet{hk14} \\
RAP & 14 &  7.40$\pm$0.19 & 3.25$\pm$0.76 &
       \citet{hk14} \\
RA  & 25 & 8.02$\pm$0.15 & 3.46$\pm$0.61 &
        \citet{ws13} \\
\hline
\end{tabular}\\
Notice: The first column shows what objects are used: QG for 
Quiescent galaxies, RA for reverberation mapped AGNs, RAC for 
reverberation mapped AGNs with classical bulges, RAP for 
reverberation mapped AGNs with pseudobulges. The second column
shows number of the used objects. The fifth column 
shows the corresponding reference.
\end{table}

\subsection{Outliers in the \msig space could be treated as clues  
to support central TDEs?}

   If $\sigma\sim81{\rm km/s}$ is taken as the best 
estimate for the velocity dispersion, it is now very interesting to 
consider a question why is SDSS J0159 an outlier in the \msig 
space? We try to answer the question after considerations of 
properties of CLAGNs and central TDEs, in order to find further 
clues on whether outliers in the \msig space could be treated 
as interesting clues to support central TDEs.

    First and foremost, through normal variabilities of accretion flows, 
similar as accretion variabilities in normal AGNs, applied to explain 
the observed properties of CLAGN SDSS J0159, there were few effects 
on dynamical structures of broad emission line clouds. Lower (stronger) 
ionization emissions lead to smaller (deeper) ionization boundary 
(smaller BLRs sizes), and then wider (narrower) broad line widths. 
It is hard to expect significantly over-estimated or under-estimated 
virial BH masses. Meanwhile, if combining with dust obscuration 
which have been applied to partly explain some CLAGNs, continuum 
(or broad line) luminosity from observed spectrum could 
be smaller than intrinsic value, leading to a bit smaller BLRs 
sizes through the R-L empirical relation. Furthermore, 
if BH mass is small as inferred from $\sigma\sim81{\rm km/s}$ through \msig 
relation for SDSS J1059 leading to much larger Eddington ratio of about 
3.6, the size of BLRs estimated through the R-L relation could be 
overestimated \citep{du18} by a factor 
about 2. It is clear that the effects of high Eddington ratio can not 
still explain the overestimated virial BH mass in SDSS J0159. Therefore, 
under the considerations of normal variations of accretion rates 
for CLAGNs combining with dust obscuration (similar as the cases on 
accretion variabilities in normal AGNs), not over-estimated but 
under-estimated virial BH mass could be expected to some extent. 
The results are not coincident with the over-estimated virial BH 
mass in SDSS J0159.

    Besides, through a central TDE applied to explain the observed 
properties of SDSS J0159, the broad emission lines were totally or 
partly from material clouds related to stellar debris. 
If the BLRs consist of materials from stellar debris, 
the BLRs may not be fully virialized. Therefore, virial method should be  
be not so efficient to estimate virial BH mass. Based on the 
simulated results on TDEs in \citet{gm14}, the length scale of stellar 
debris $R_{\rm TDE}$ (the parameter $r_o$ in \citet{gm14}) could be 
about $r\sim t^{2/3}$, however the time-dependent accretion luminosity 
from the central TDE is decreasing after the time of peak accretion 
($t_{\rm p}$). Therefore, $L_{\rm TDE}$ and $R_{\rm TDE}$ can not 
follow the empirical R-L relation determined from normal reverberation 
mapped AGNs. 

\begin{figure}
\centering\includegraphics[width = 8cm,height=5cm]{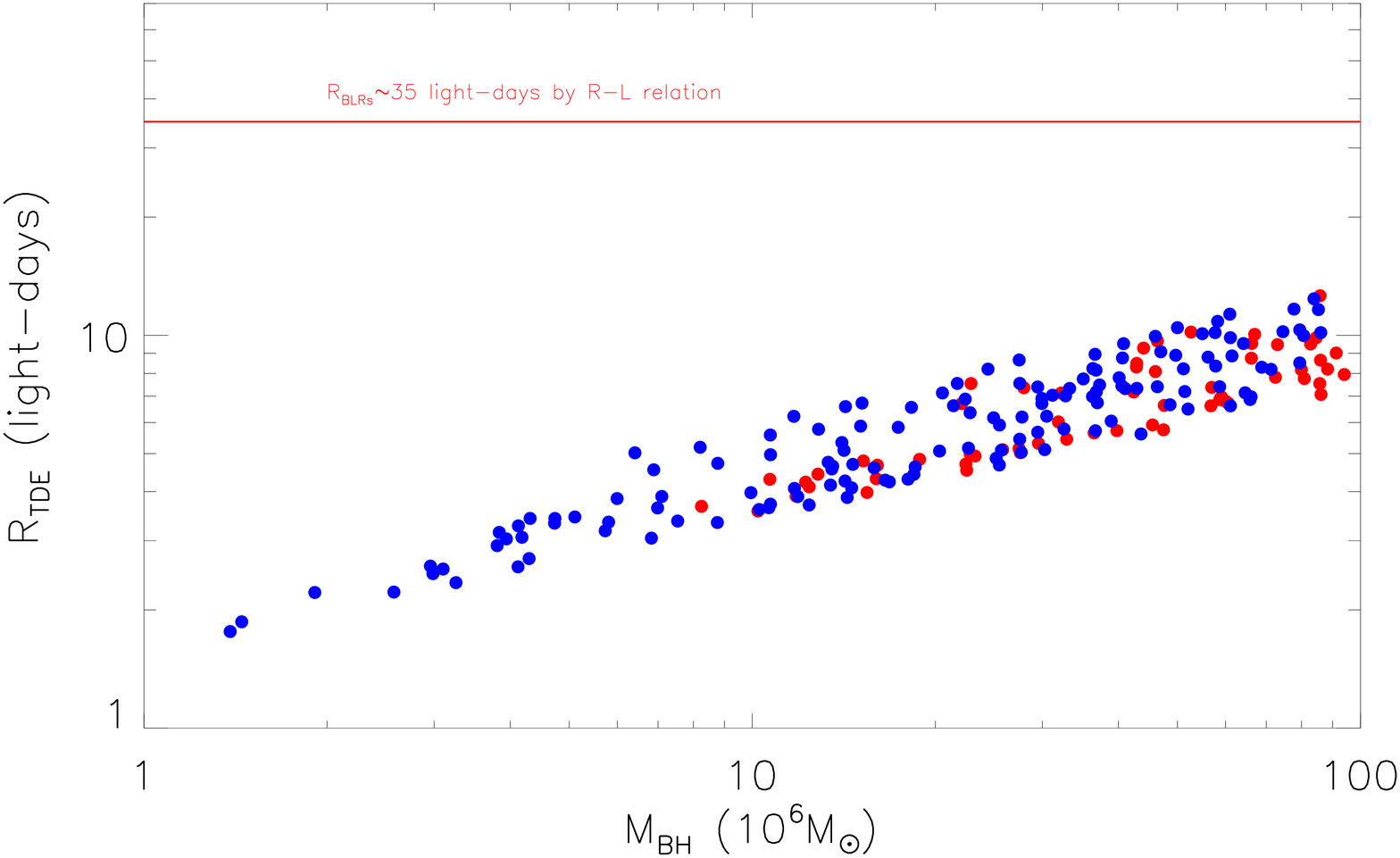}
\caption{Properties of $R_{\rm TDE}$ at $t_{\rm p}$ calculated 
through TDE simulations. Circles in red and in blue represent 
the results for the cases with central tidal disrupted stars with 
polytropic indices of $4/3$ and $5/3$, respectively. 
The horizontal red line shows the position of BLRs sizes 
$R_{\rm BLRs}~\sim~35~{\rm light-days}$ determined by the 
R-L relation for SDSS J0159 in 2000.
}
\label{rtde}
\end{figure}

     In order to show much clear results on $R_{\rm TDE}$, 
Fig.~\ref{rtde} shows the calculated $R_{\rm TDE}$ at $t_{\rm p}$ 
from TDEs with central BH masses randomly collected from 
$10^6~{\rm M_\odot}$ to $10^8~{\rm M_\odot}$, central 
tidal disrupted stars randomly collected with masses from 
$0.1~{\rm M_\odot}$ to $20~{\rm M_\odot}$ and with polytropic 
indices of $4/3$ and $5/3$, respectively. Here, the following four 
criteria have been accepted. The first is that the calculated 
tidal disruption radius must be larger than event horizon of 
central BH. The second is that the calculated peak accretion 
rate from TDEs should be from $0.2{\rm M_\odot/yr}$ to 
$0.3{\rm M_\odot/yr}$, leading to peak bolometric luminosity 
from TDEs around $1.5\pm0.2\times10^{45}~{\rm erg/s}$ (the value 
for SDSS J0159 in 2000 with 10-15 as the ratio of bolometric 
luminosity to optical luminosity). The third is that the 
calculated $t_{\rm p}$ should be from 0.8yrs to 2yrs, similar 
as the case in SDSS J0159 as discussed in \citet{md15}. The 
forth is the mass-radius relation has been accepted to the main 
sequence stars, but with maximum uncertainties of about 50\%. 
For SDSS J0159, the estimated BLRs size in 2000 (around $t_{\rm p}$ 
for the TDE in SDSS J0159) is about 35 light-days by the R-L 
relation, much larger than the calculated $R_{\rm TDE}$ shown 
in Fig.~\ref{rtde}. Therefore, effects of central TDEs should 
lead to over-estimated virial BH masses through properties of 
single-epoch spectrum observed around $t_{\rm p}$ due to larger 
BLRs sizes than intrinsic values, if there were 
strong contributions of stellar debris to broad line regions.

   Last but not the least, we give some further discussions 
on the case for the central TDE in SDSS J0159. Based on the 
measured stellar velocity dispersion, the BH mass can be 
roughly estimated as $2.15\times10^6~{\rm M_\odot}$ through 
the \msig relation in equation (3), indicating $R_{\rm TDE}$ 
about 2~light-days by the results shown in Fig.~\ref{rtde}, 
about 17 times smaller than the luminosity expected BLRs 
sizes by the R-L relation. Actually, $R_{\rm TDE}$ represents 
the outer boundaries of materials in disk-like regions in 
TDEs, the velocity-dependent locations of emission line clouds 
should be much smaller than $R_{\rm TDE}$. Thus, the calculated 
virial BH mass could be several tens of times larger 
than intrinsic BH mass for SDSS J0159, similar as the results 
shown in Fig.~\ref{msig}, if accepted that BH mass through the 
\msig relation reliable enough. In other words, the 
over-estimated virial BH mass leading to SDSS J0159 as 
an outlier in the \msig space could be treated as a strong 
evidence to support a central TDE in SDSS J0159. 

   In brief, after considerations of contributions of stellar 
debris from central TDEs to broad line clouds, it is inefficient 
to estimate BLRs sizes by the commonly applied R-L empirical 
relation, leading to over-estimated virial BH masses measured 
through observed broad line properties of line width and line 
(and/or continuum) luminosity. Hence, over-estimated virial BH 
masses in the \msig space could be treated as interesting clues 
to support central TDEs in AGNs.

\section{Conclusions}

   Finally, we give our main conclusions as follows. 
Based on different methods applied to describe the 
spectral absorption features, the stellar velocity dispersion 
can be well measured as $\sigma\sim81~{\rm km/s}$ for 
the interesting object SDSS J0159, a CLAGN and also a 
host galaxy of a central TDE. Comparing to the reported 
virial BH mass of $\sim10^8~{\rm M_\odot}$ in the 
literature, the small $\sigma$ leads SDSS J0159 to be 
an apparent outlier in the well-known \msig space. After 
considerations of contributions of TDEs to observed 
broad line emission clouds, the over-estimated virial BH 
mass can be well expected, due to luminosity dependent 
BLRs sizes through the R-L relation much larger than 
the true intrinsic sizes of broad line emission clouds 
from the central TDE. Therefore, the over-estimated virial 
BH mass estimated through line width and line luminosity 
in the \msig space could be treated as interesting 
clues to support central TDEs.

\section*{Acknowledgements}
Zhang, Bao \& Yuan gratefully acknowledge the anonymous 
referee for giving us constructive comments and suggestions to greatly 
improve our paper. Zhang gratefully acknowledges the kind 
support of Starting Research Fund of Nanjing Normal University. 
This Letter has made use of the data from the 
SDSS projects. The SDSS-III web site is http://www.sdss3.org/. 
SDSS-III is managed by the Astrophysical 
Research Consortium for the Participating Institutions of the 
SDSS-III Collaboration including University of Arizona, Brazilian 
Participation Group, Brookhaven National Laboratory, Carnegie Mellon 
University, University of Florida, French Participation Group, 
German Participation Group, Harvard University, Instituto de 
Astrofisica de Canarias, Michigan State/Notre Dame/JINA Participation 
Group, Johns Hopkins University, Lawrence Berkeley National 
Laboratory, Max Planck Institute for Astrophysics, Max Planck 
Institute for Extraterrestrial Physics, New Mexico State University, 
New York University, Ohio State University, Pennsylvania State 
University, University of Portsmouth, Princeton University, Spanish 
Participation Group, University of Tokyo, University of Utah, 
Vanderbilt University, University of Virginia, University of 
Washington, and Yale University.

\label{lastpage}
\end{document}